\documentclass[10 pt, conference]{IEEEtran}
\usepackage[utf8x]{inputenc}
\usepackage{balance}
\usepackage{comment}
\usepackage{color, soul}
\usepackage{graphicx}
\usepackage{textcomp}
\usepackage{adjustbox}
\usepackage{array}
\usepackage{booktabs}
\usepackage{multirow}
\usepackage{tikz}
\usepackage{subfig}
\usepackage{ulem}
\usepackage{url}

\usepackage[final]{microtype}

\IEEEoverridecommandlockouts                              
\title{\LARGE \bf
Technical Report: Gone in 20 Seconds - Overview of a Password Vulnerability in Siemens HMIs
}

\author{Joseph Gardiner and Awais Rashid \\ Bristol Cyber Security Group, University of Bristol \\ Bristol, UK \\\{joe.gardiner,awais.rashid\}@bristol.ac.uk}

\begin{document}

\maketitle
\thispagestyle{empty}
\pagestyle{empty}

\begin{abstract}

Siemens produce a range of industrial human machine interface (HMI) screens which allow operators to both view information about and control physical processes. For scenarios where an operator cannot physically access the screen, Siemens provide the SM@rtServer features on HMIs, which when activated provides remote access either through their own Sm@rtClient application, or through third party VNC client software. 

Through analysing this server, we discovered a lack of protection against brute-force password attacks on basic devices. On advanced devices which include a brute-force protection mechanism, we discovered an attacker strategy that is able to evade the mechanism allowing for unlimited password guess attempts with minimal effect on the guess rate. This vulnerability has been assigned two CVEs - CVE-2020-15786 and CVE-2020-157867.

In this report, we provide an overview of this vulnerability, discuss the impact of a successful exploitation and propose mitigations to provide protection against this vulnerability. 

This report accompanies a demo presented at CPSIoTSec 2020~\cite{hmi:cpsiotsec}.
\end{abstract}

\section{Introduction}

Industrial control systems (ICS) are the systems responsible for the control and  operation of both critical national infrastructure (CNI), including oil and gas, water treatment and power generation, as well as manufacturing processes. Industrial control systems are made up of many speciality devices, including programmable logic controllers (PLCs), remote telemetry units (RTUs) and human-machine interfaces (HMIs), with major manufacturers including Siemens, Allen Bradley, Honeywell, Schneider Electric and General Electric. These systems are often referred to as Operational Technology (OT). In ICS, safety is the number one concern, with devices designed to operate reliably for many years. The security of such devices was largely physical - they were designed to sit without an internet connection behind locked doors. In modern times, however, this is not the case with devices regularly being connected to the internet. Incidents such as the Stuxnet and Triton malware, which specifically target industrial systems, and legislation such as the European Network and Information Systems (NIS) directive have put the cyber security of industrial systems very much in the focus. As part of this, there are an increasing number of vulnerabilities being discovered, and eventually patched, in industrial devices.  

Human machine interfaces (HMIs) primarily refer to a physical device which is designed to be installed in physical proximity to a physical process. HMI screens are programmed to both provide a display of information relating to the physical process below, as well as allow operators to provide inputs to the control system to control and manage physical processes. These screens can vary from a few inches in size up to ``full size'' monitors, with modern devices usually featuring a touchscreen, and in some cases a set of physical inputs including buttons and dials. Most ICS manufacturers produce some range of HMI screens, including Siemens who produce a wide range of these devices, two of which can be seen in Figure~\ref{fig:hmis}.

Some HMIs support remote access, which allows operators in a central location to access screens that human operators are unable to access. This provides obvious benefits, allowing engineers to correct issues remotely as well as monitor and control processes. Device manufacturers provide their own methods for this remote access. Communication is usually achieved over network connections, with most modern devices featuring an ethernet port and/or wireless connectivity. In the case of Siemens, the primary method is through the use of the VNC-based Sm@rtServer system available on most of their HMI range, which provides access through a Sm@rtClient application (available for PC, Android and iOS), as well as through third party VNC clients. 

We discovered a vulnerability in Siemens HMI products that allows an attacker to be able to brute force the Sm@rtServer password. On basic devices, we find that there is no protection against brute forcing the Sm@rtServer, allowing for the use of existing online password cracking tools. We discover that on higher end devices, the Sm@rtServer employs a form of brute-force prevention, which we were able to evade allowing for slightly slower, but still overall successful, brute force attempts. Successfully guessing this password could in some cases grant an attacker full control over the HMI screen, and therefore control over the underlying process, causing a potentially dangerous, life threatening situation. 

The remainder of this paper is structured as follows: In section~\ref{sec:background} we provide an overview of the Siemens HMI range and Sm@rtServer service. In Section~\ref{sec:vulnerability} we provide an overview of the vulnerability and how it was discovered, and discuss the potential impact of successful exploitation in Section~\ref{sec:impact}. Finally, we propose some mitigations in Section~\ref{sec:solutions}.

\subsection{Responsible Disclosure}

The discussed vulnerabilities were reported to Siemens on the 13th of February 2020. Siemens have accepted the vulnerabilities and have assigned two CVEs to this vulnerability:
\begin{itemize}
    \item \textbf{CVE-2020-15786} for the brute forcing of passwords.
    \item \textbf{CVE-2020-15787} for the issues of password truncation when using a VNC client. 
\end{itemize}
The vulnerability is addressed in Siemens Security Advisory (SSA) 524525~\cite{ssahmi}. Whilst they are preparing updates to counter the problem, the suggested countermeasure is to apply ``Defence-in-Depth''~\cite{siemensoperationalhuidelines}.
\section{Background}\label{sec:background}

\subsection{Siemens HMI Range}

\begin{figure}
\subfloat[A Siemens KTP700 Basic HMI panel (mounted onto box), controlling a virtual process \label{fig:ktp700}]{\includegraphics[width=1\columnwidth]{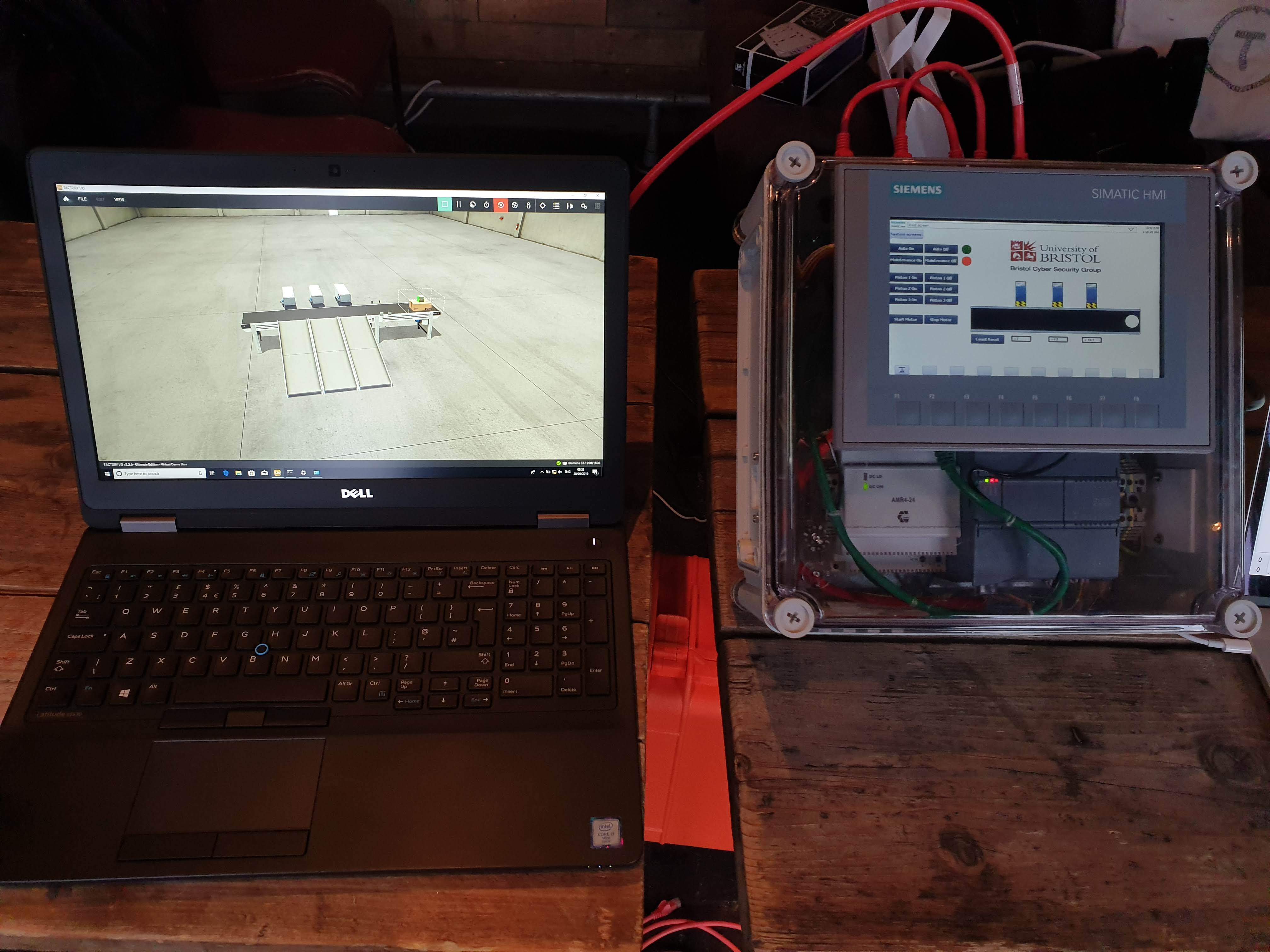}}\\
\subfloat[A Siemens Mobile Panel 277 IWLAN V2, mounted on cabinet\label{fig:wireless}]{\includegraphics[width=1\columnwidth]{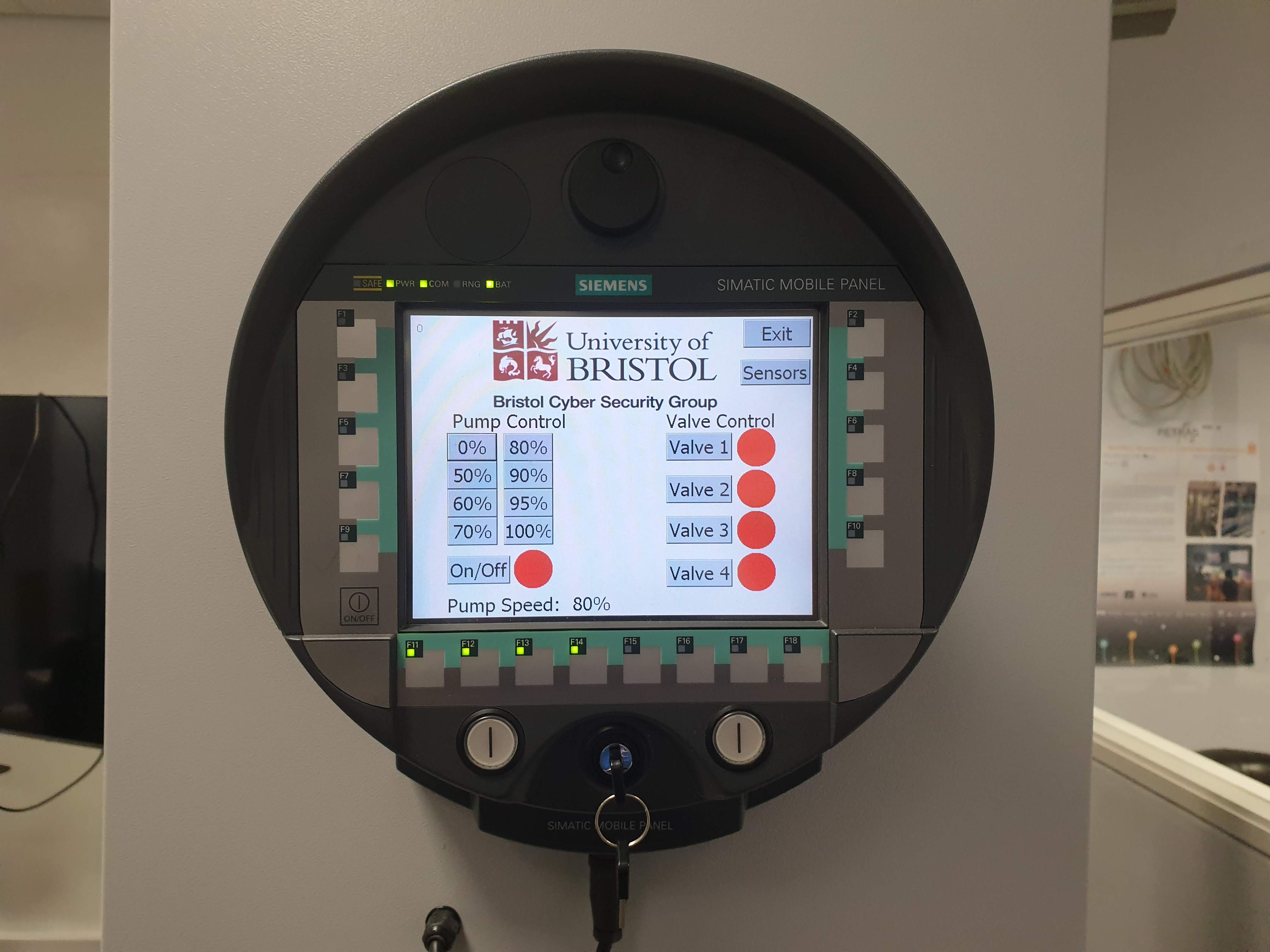}}
\caption{Two Siemens HMI devices}
\label{fig:hmis}
\end{figure}
The Siemens HMI range is currently split into two main branches: the ``Basic'' range and the ``Advanced'' range. 

The Basic range are simple panels, which feature a touchscreen and buttons, available in 4, 7, 9 and 12-inch models, with the KTP product name. A KTP700 7-inch display is visible in Figure~\ref{fig:ktp700}. Basic panels feature a custom Linux-based operating system.

The Advanced range includes the comfort panel range, which are available in sizes from 7 to 22 inches, both in touchscreen and non-touchscreen variants. These devices utilise Windows CE as the base operating system.

The advanced range also includes mobile panels. Part of this range are similar to the basic range, but rather than requiring mounting are self-contained, wired units that can be moved around. Also available are a series of wireless panels which use WiFi connections to communicate with other devices, such as the Mobile Panel 277 IWLAN, as seen in Figure~\ref{fig:wireless}. These wireless devices also use Windows CE as the base operating system.

All of the mentioned devices utilise a network connection to communicate with other devices, including PLCs. The Siemens PROFINET protocol is used over ethernet or wireless for this purpose. The network connection is also used for configuring and programming the devices using the Siemens TIA portal software. 

\subsubsection{Past vulnerabilities}

There have been a number of vulnerabilities that have been found in Siemens HMI products, and we will discuss a small number of recent vulnerabilities here.  For example, CVE-2020-7592 describes unencrypted communication between HMI screens and the configuration software, which could allow an attacker with network access to potentially capture sensitive information. CVE-2019-10926 is a vulnerability affecting many PROFINET devices, where a large number of specially-crafted UDP packets can trigger a denial-of-service condition. Finally, CVE-2019-6577 is one of many vulnerabilities affecting the web server included on some HMI devices, in this case a cross-site scripting attack is possible if an attacker is able to modify particular parts of the device configuration via SNMP.

\subsection{Siemens Sm@rtServer and Sm@rtClient}
The Siemens Sm@rtServer is primarily marketed as a solution for the remote servicing of plants, in particular for handling problems that cannot be solved by staff on site~\cite{smartserverscenarios}. Primarily, the Sm@rtServer application provides remote access to an HMI panel, allowing a remote operator full control over an HMI, allowing them to both read the state of processes and issue commands to the system. 

The primary way to access the Sm@rtServer is through the Sm@rtClient application, available for Windows, Android and iOS. An example of a connection to a KTP700 basic panel can be seen in Figure~\ref{fig:basicconfig}, and a TP1500 Comfort panel in Figure~\ref{fig:TP1500home}. The Sm@rtClient software allows interaction both with the screen, and a mapping of the physical inputs (such as buttons and dials) that feature on many HMIs. The Sm@rtServer also supports connections by third party VNC clients.

Through examining the communication between the Sm@rtClient software and Sm@rtServer, we identify that TightVNC is used as the underlying VNC protocol, utilising the ``Tight'' security type\footnote{\url{https://vncdotool.readthedocs.io/en/0.8.0/rfbproto.html#tight-security-type}}. For third party VNC clients, standard VNC authentication is used, limiting password lengths to 8 character. When using the Sm@rtClient, an alternate authentication scheme is used which permits passwords greater than 8 characters, which we believe is custom to Siemens. When examining the traffic, Tight authentication capabilities of ``SICRSCHANNEL'' and ``SICRSCHANLPW'' are offered, the second of which is selected by Sm@rtClient, and both of which are rejected by other VNC clients. We were unable to find any other references to this through web searches, which leads us to believe this is a Siemens defined authentication capability that permits longer passwords.

Whilst the Sm@rtServer is installed by default on all tested HMIs, it requires a licence to operate fully. The server can, however, run without a license with full functionality, but a dialog appears frequently on the screen advising that the running server is unlicensed. 

We have observed two versions of the Sm@rtServer system on different HMIs. A basic, and more vulnerable, version is available on the KTP range of basic panels, with fewer options that can be configured. On higher end panels with a Windows CE underlying OS, such as the Comfort and Mobile range, a more advanced version of the Sm@rtServer is used with further configuration options. 

By default, traffic from the Sm@rtClient app to the Sm@rtServer is unencrypted, though encryption is available on the full version (though this requires activation). Connections from third party VNC clients are always unencrypted (except for password transmission). Whilst we do not explore this issue in this paper, unencrypted VNC communication could potentially leak highly sensitive information about the industrial process.

\subsection{Setup and configuration}
\begin{figure*}[!ht]
    \centering
      \subfloat[
      Basic Sm@rtServer configuration\label{fig:basicconfig}]{\includegraphics[width=0.32\textwidth]{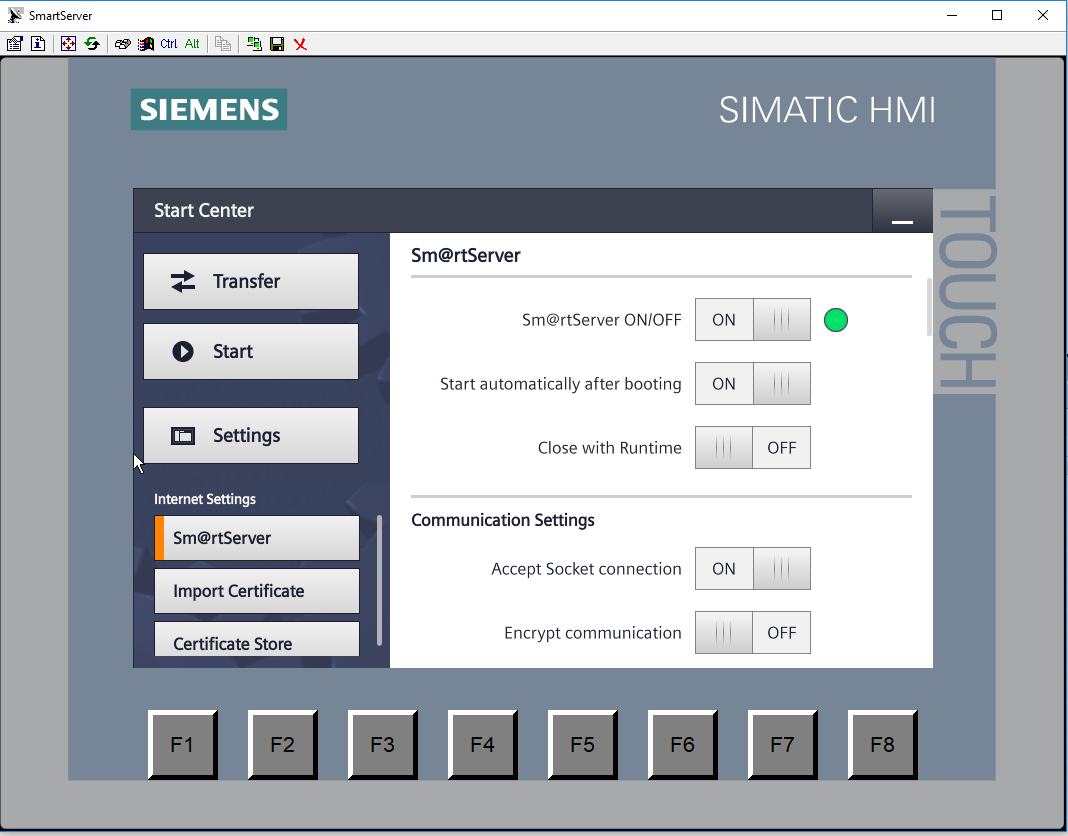}}
        \subfloat[
        Advanced Sm@rtServer configuration\label{fig:smartserverconfig}]{\includegraphics[width=0.32\textwidth]{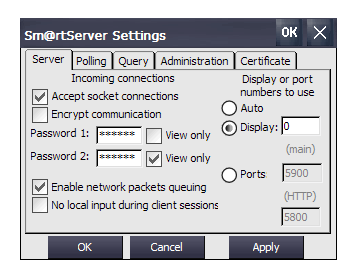}}
        \subfloat[
        Advanced Sm@rtServer password length warning\label{fig:smartserverwarning}]{    \includegraphics[width=0.32\textwidth]{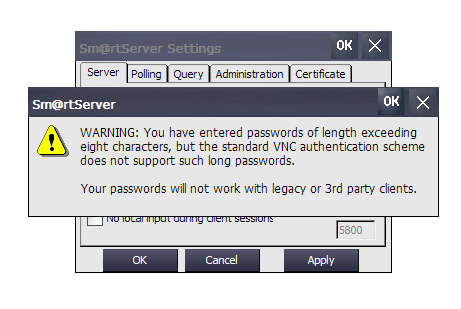}
}   
    \caption{ Sm@rtServer configuration}

    \label{fig:subfigname}
\end{figure*}

The Sm@rtServer is activated on HMI screens through the Siemens TIA portal software. However, this only activates the functionality on the device. Configuring, including setting an access password, and starting the server is done within the settings screens of the device itself. Figure~\ref{fig:basicconfig} shows an example of this configuration screen on the Basic range of HMIs, and Figure~\ref{fig:smartserverconfig} shows the main configuration screen for the advanced ranges of HMIs including the Comfort and Wireless ranges.  

Once the server is activated, the primary task is to set a password. This has to be done on the screen itself. On both versions of the Sm@rtServer, two passwords can be set. By default, the second is set to allow read-only access to the screen, though this option can be selected for either password. Due to the limitation of many VNC clients, if a password longer than eight characters is entered, a warning, as seen in Figure~\ref{fig:smartserverwarning}, is displayed. This advises that the password will not work with legacy or third party clients, which is true. The user can however ignore this warning and set a longer password, which will limit them to only use the official Sm@rtClient client software. This message may encourage users to select passwords that are 8 characters or shorter to allow for future access via third party VNC clients.  There is no form of strength meter or policy on the quality of the set password, which may lead users to enter a less secure password as has been demonstrated in user studies~\cite{10.5555/2362793.2362798}. Further, producing warning when setting a longer password for those who not require VNC access and can benefit from the enhanced security of a longer password is counter-productive.

There is also concern of the security of passwords that are entered on the devices directly. Often, the HMI touchscreens are far less precise than most modern touchscreens, such as those found on smartphones and tablets, and typing on them required use of a clunky on screen keyboard. As the password is set on the HMI directly, this may lead the operator to set a less secure password. In a study by Melicher et. al~\cite{10.1145/2858036.2858384}, they find that creating passwords on mobile devices such as phones and tablets is more frustrating and time consuming for users, and users set less secure passwords with fewer capital letters and grouped special characters. This confirms similar results from past studies~\cite{10.1145/2639189.2639218}.

\section{Vulnerability}\label{sec:vulnerability}
The vulnerability takes advantage of the Sm@rtServers support of the VNC protocol, allowing it to be used with most VNC clients. By allowing users to connect using VNC clients, there is an opportunity for an attacker to brute force the password using off-the-shelf tools for basic panels, and through scripting for advanced panels, such as the Comfort range. 

Through breaking the password, the attacker can gain control of the HMI screen. Whilst they have full control of the screen itself over a VNC connection, a connection over VNC does not allow the attacker to control the physical inputs (such as buttons and dials) that feature on many HMIs. If the attacker wishes to do this, they would need to use the official Sm@rtClient application with the stolen password, which is able to map these inputs. This application is free to download from the Siemens website and so this is not a barrier for an attacker.

\subsection{Preconditions}
In order to exploit the vulnerability, an attacker would need network access to the device. This could either be through access to an OT network through compromise, or through devices being connected directly to the internet. A simple search on Shodan for devices with port 102 open (the port used by the Siemens S7 protocol used by devices) reveals a large number of Siemens devices publicly accessible\footnote{\url{https://www.shodan.io/search?query=port\%3A102}}.

The Sm@rtServer feature, whilst available on all devices, is not activated by default. This means that only devices for which the service has been activated, and configured, are vulnerable to the attack. 

The impact of the attack can vary depending on how the HMI screens have been designed by the operator. We discuss the impact of the attack more in Section~\ref{sec:impact}.

\begin{table}[]
\scalebox{1}{
\begin{tabular}{|l|l|}
\hline
{\textbf{Device}} & {\textbf{Firmware Version}}  \\ \hline
KTP700 Basic                      &   15.01.00.00\_24.01   \\ \hline
TP1500 Comfort Panel V2     &    V15.01.00.00\_26.01             \\ \hline
  MobilePanel 277 IWLAN V2                    &  12.0.0.0                              \\ \hline

\end{tabular}}
\caption{Tested devices and firmware versions}
\label{tab:devices}
\vspace{-18pt}
\end{table}
\subsection{Basic panels}

For basic panels, there is no form of restriction on login attempts when connecting over VNC. This means that we can use any standard tool capable of cracking VNC passwords, such as Hydra~\footnote{\url{https://github.com/vanhauser-thc/thc-hydra}}. 

When using Hydra, the attacker simply needs to specify the VNC protocol, state that no username is required and provide a password list. Hydra recommends that for VNC brute forcing, no more than 4 threads are used to prevent overloading the server. 

Hydra will not open a connection to the HMI once a correct password is found, and will only print to the screen. The attacker can configure Hydra to stop guessing after the first successful password is found, or to keep guessing. As we discuss in section~\ref{sec:passwordlimit}, this is worth doing as there may be multiple passwords that are valid.

\subsection{Advanced panels}
On the Comfort and Mobile Wireless HMIs, there is some level of protection against brute forcing of the password. If an attacker simply tries and guesses passwords using the same technique as the basic panel, the connection is refused after 5 incorrect guesses have been made. This refused connection is different to the normal authentication error when attempting with an incorrect password. 

We first attempted to evade this limitation through slowing down the guess rate. However, we found that even with a very slow guess rate of 1 every 10 seconds, the 5 guess limitation still applies. We then attempted to make 5 guesses at full speed, then waiting for increasing time periods before making further guesses, which still provided a negative result. 

The next step was to change IP addresses after 5 guesses, which has the effect of a ``new'' device making guesses. We did this by simply changing the IP of the attacker machine manually after every 5 guesses by switching network profiles within the Kali VM (we used a slow guess rate of 1 per second to be able to achieve this switch). After the 5 guesses with the first IP, we switched to the second and were successfully able to make a further 5 guesses before being locked out for the second IP. 

This made it clear that the protection mechanism works on an IP basis, meaning that, as long as the IP changes, one can make further guesses. Our first assumption was that the device maintains a single register which will store a single IP and its failed guess count, meaning that one can switch between 2 IP addresses every 5 guesses and make an unlimited number of guesses. 

We first tested this theory with 2 IP addresses, making 5 guesses at a rate of 1 per second before switching between them. We found that on returning to the first IP, the device rejected the connection. We then gradually increased the delay per guess to see if there is a time factor involved, and at 1 guess per every 4 seconds, we were able to continuously make guesses without a rejected connection. This implies that there is a 20 second lockout period. 

We again tested with a single IP with a 20 second sleep period after 5 guesses, which still resulted in rejected connections. We eventually concluded that in order to make continuous attempts, the attacker needs to switch IP addresses every 5 guesses and not return to an IP address within 20 seconds. Therefore, if the attacker can increase the pool of IP addresses they can use, they can increase the rate of guesses. 

We produced a bash script for automating this process, which modified the host IP address automatically. We tested with up to 8 IP addresses, and achieved a rate of 1 guess every 0.5 seconds. If the attacker is able to use the full 254 IPs available in a single subnet, they would be able to make approximately 3800 guesses per minute.

\subsubsection{Limitations}
Obviously, changing the IP address can have an effect on the host machine. If the ``attacker'' machine was a compromised host within the network, then changing the IP could cause services already running on that machine to fail, alerting administrators. Similarly, increasing the count of IPs makes the attack more noticeable, especially at high guess rates. For an attacker who wishes to remain covert, the best approach would be to use 2 IP addresses at the slowest guess rate possible. The brute force will take a long time but will result in minimal additional packets on the network. 

An attacker who has control of multiple machines could also split the load across these machines rather than modify the IP address of a single machine, which would prevent the issue of damaging existing services. A botnet could provide an easy method of achieving a large pool of machines. 

\begin{table}[]
\scalebox{1}{
\begin{tabular}{|p{2.5cm}|p{2.5cm}|p{2cm}|}
\hline
{\textbf{Number of IP addresses}} & { \textbf{Time per guess (seconds)}} & { \textbf{Guesses per minute}} \\ \hline
2                                     & 4                                       & 15                                \\ \hline
4                                     & 1                                       & 60                                \\ \hline
8                                     & 0.5                                     & 120                               \\ \hline
256                                   & 0.015625                                & 3840                              \\ \hline
\end{tabular}}
\caption{Guess rate by number of IPs}
\label{tab:guesses}
\vspace{-12pt}
\end{table}

\begin{figure*}[!ht]
    \centering
        \subfloat[
        TP1500 Home Screen\label{fig:TP1500home}]{\includegraphics[width=0.5\textwidth]{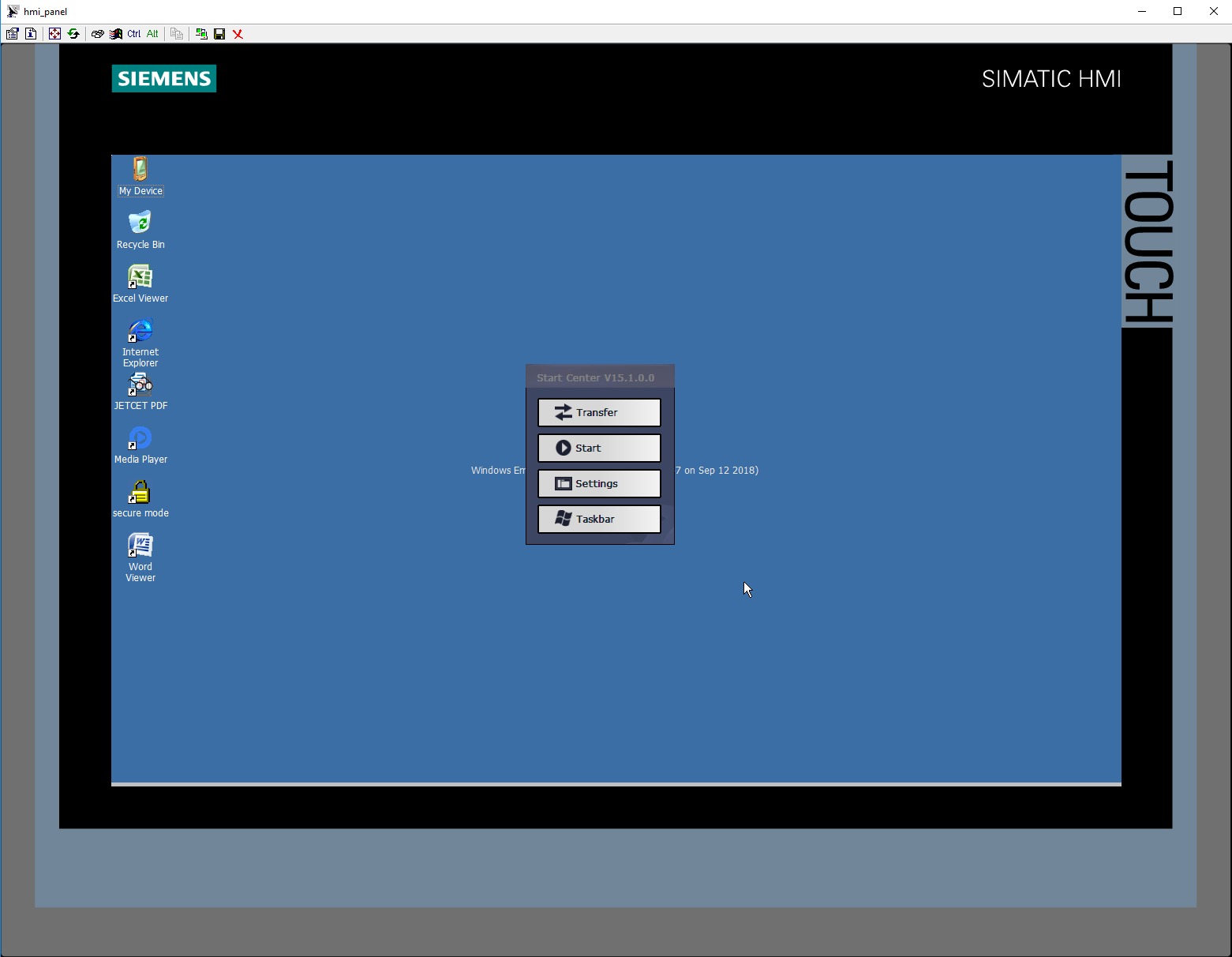}}
        \subfloat[
        TP1500 Control Panel\label{fig:TP1500control}]{\includegraphics[width=0.5\textwidth]{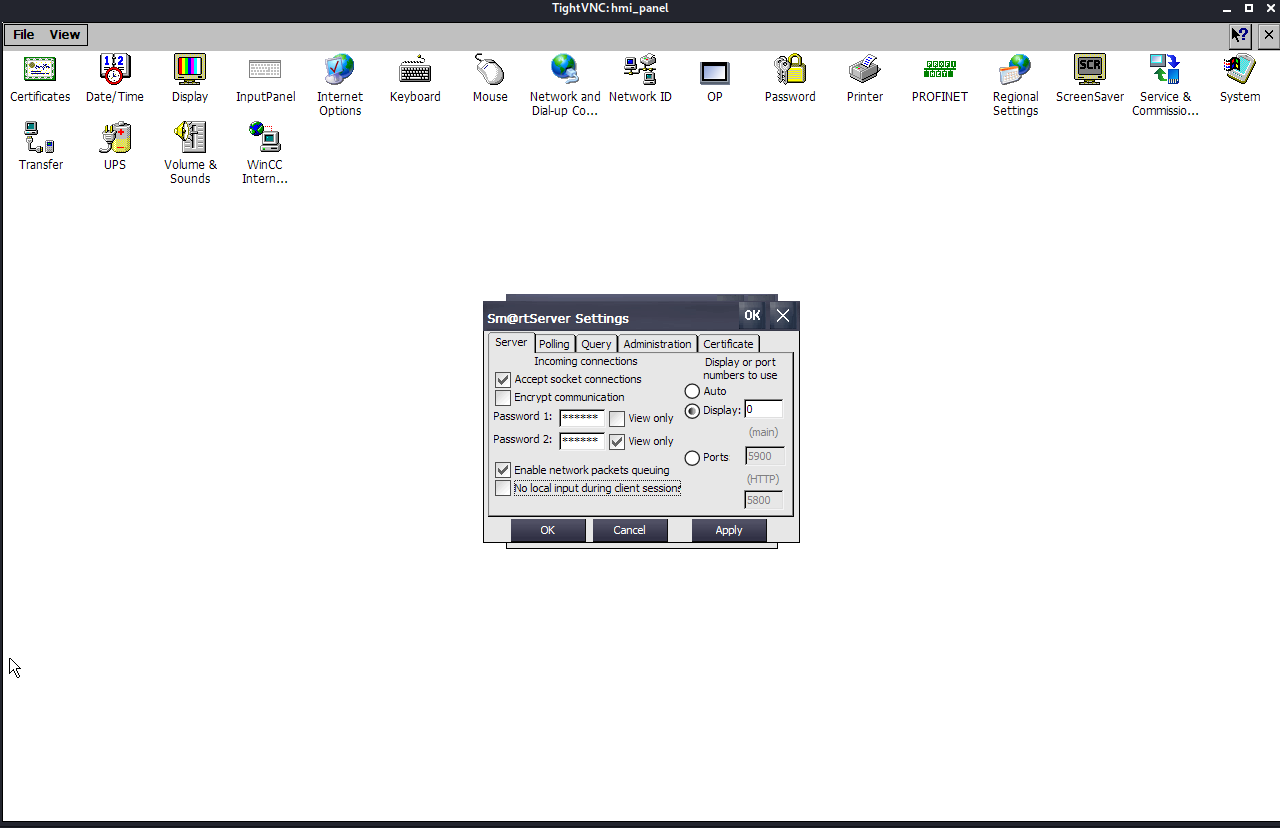}
}
    \caption{TP1500 Comfort home screen and control panel}

    \label{fig:subfigname}
\end{figure*}

\subsection{Limitations of VNC passwords}\label{sec:passwordlimit}
In most versions of VNC, passwords are limited to 8 characters. In some cases, this results in passwords longer than 8 characters being truncated by VNC clients. As an example, if a Sm@rtServer password longer than 8 characters is set, most third party VNC clients will fail to authenticate. When setting a longer password, the HMI displays a warning to the user (such as in Figure~\ref{fig:smartserverwarning}).

One effect of this is that, when brute forcing passwords using a word list and a VNC-based client, an attacker can find ``successful'' passwords that are longer than 8 characters, as long as an 8 character password has been set. For example, if the password is set to ``password'', then guesses such as ``password1'' or ``password!'' are marked as successful, and will allow authentication through VNC clients. This can potentially make brute-force attempts easier. As an example, the wordlist that we used for testing featured ``password!'' and ``password123'' above ``password'' on the list, and so we would get a successful password that will allow access earlier than we otherwise would.

\section{Impact}\label{sec:impact}

The primary impact of this vulnerability is that if an attacker gains control of the HMI screen, they can potentially control the underlying physical process. This is entirely dependent on the design of the HMI screens programmed by the operator. In some cases this screen can provide full control over the process, in others it will only display information about the running process. Similarly, the Sm@rtServer can be configured to provide read only access to the device. Even in the scenario where an attacker cannot control the process directly, they can gain valuable information about the process that can assist in carrying out further attacks against the associated devices~\cite{1d347704cb6f4f6e87badf28df6158e4,186136471bc949e09d41ac61539db9da}.

It is worth noting that Siemens, in their ``Security Guidelines for SIMATIC HMI Operator Panels'' document~\cite{hmisecurity}, suggest limiting direct control of physical processes to remote users. This does however require further programming effort on behalf of the operator.

If the programmed HMI screen features an exit button (as is the default on screen templates), which is not protected by a local account on the device (distinct from the Sm@rtServer authentication), the attacker can exit to the configuration screens for the device itself. This can have three affects. First, this would allow an attacker to change the configuration of a device, for example, changing the assigned IP address. On Windows CE based devices this includes access to the control panel, as seen in Figure~\ref{fig:TP1500control}.

Secondly, further applications are available on the Windows based devices. These can include Internet Explorer, Word and Excel Viewer, and a command prompt access, which could all be used to perform covert attacks against other devices from a trusted device, avoiding firewalls. 

Finally, as the attacker can gain access to the Sm@rtServer configuration screen, they can modify any settings relating to the screen. This can include changing the configured password. The most notable effect, however, is that they can configure the Sm@rtServer to disable the physical touchscreen whilst a remote connection is active. This setting can be seen in Figure~\ref{fig:smartserverconfig}. This would mean that an attacker would have full control over the screen, whilst a local operator would not be able to take back control until the attacker has closed the connection.

\section{Solutions}\label{sec:solutions}

We will now propose mitigations for the immediate protections for devices, as well as longer term solutions. Note that this is in additional to the recommendation from Siemens in their security advisory relating to this vulnerability to apply ``Defence-in-Depth''~\cite{siemensoperationalhuidelines}. 

\subsection{Mitigation}
The primary mitigation to prevent this attack is to disable the Sm@rtServer feature if possible until a software update has been released. However if the Sm@rtServer is required to maintain plant operation, there are other measures than can be taken to provide short-term mitigation against the attack:
\begin{itemize}
    \item Limit client use to the Sm@rtClient application only, which allows for password lengths greater than 8 characters, and ensure standard good password practice is maintained when choosing the required password. If VNC access is required, then ensure a strong 8 character password is chosen to make brute-forcing as difficult as possible. 
    \item Ensure that no two HMIs share the same password. 
    \item Ensure that the Sm@rtServer is configured to allow read-only access if control input is not required. 
    \item Configure firewall rules to only allow specific client IP addresses to access target devices over VNC connections. 
    \item Configure intrusion detection systems to look for large volumes of failed VNC connection requests to single HMIs. 
    \item Configure screens to prevent users from exiting to the system configuration screens.
    \item Configure on-screen confirmation for remote connections (for scenarios where operators are present on site and remote access is for remote support only)~\cite{hmisecurity}.
    \item Configure a different password to prevent access to the settings menus without further authentication~\cite{hmisecurity}. 
    
\end{itemize}

\subsection{Solutions}

For a longer term solution, the device needs to reliably be able to temporarily limit password guessing attempts, but not block the client IP for long periods as this could prevent genuine access to the device. Through our interaction with Siemens whilst disclosing this vulnerability, they are intending to patch this aspect of the vulnerability. 

The system should prevent users, who are accessing over a remote connection, from being able to select the options to lock the screen whilst a remote user is connected, as this would prevent the attacker from locking the screen to local operators.

The long term solution would be to migrate away from the VNC protocol for remote access to allow other methods of connection, or if VNC protocol access is maintained, a move to a more effective authentication and authorisation mechanism that does not rely on passwords and only permits authorised clients access.

Authentication and authorisation is a widely-studied field, with an increasing number of alternative to password-based authentication methods. Many of these could be applicable to OT environments and devices and provide stronger authentication for what is often safety critical pieces of equipment. This will not be a straightforward process, however, due to the safety and real-time operation requirements of industrial devices, and usability issues with touchscreens or basic physical keyboards. 
\section{Conclusion}\label{sec:disclosure}
In this paper we have described a vulnerability that we have discovered in Siemens HMI products. The vulnerability allows an attacker to brute force the password of the Sm@rtServer remote access solution due to insufficient protections against such brute-force mechanisms. Successful exploitation of this vulnerability potentially grants an attacker great control over the underlying physical process, allowing an attacker to cause large amounts of damage to physical property, loss of critical services such as power generation or water treatment, or in extreme cases physical harm to humans. 

We also propose mitigations and push for stronger authentication of HMI panels, and other ICS devices. It is time that OT security in general draws upon the authentication and authorisation advances made in other domains, though we accept that there are safety and real-time considerations for OT devices which makes direct application of such techniques not a straight forward process. 

\textbf{Acknowledgements: }
We would like to thank Emmanouil Samanis of Bristol Cyber Security Group and Richard J. Thomas of the University of Birmingham for their assistance during the testing of this vulnerability. We would also like to thank the Siemens ProductCERT team for constructively engaging with us during the disclosure process.

\bibliographystyle{abbrv}
\bibliography{references.bib}

\end{document}